\documentstyle[12pt]{article}

\newcommand{\vev}[1]{\langle #1 \rangle}                 % VEV
\newcommand{\com}[2]{\left[ #1,#2 \right]}               % commutator
\newcommand{\resetcounter}{\setcounter{equation}{0}}     % set counter to zero
\newcommand{\lagrange}{{\cal L}}

\newcommand{\D}  {{\cal D}}               % covariant derivative
\newcommand{\EPS}{{\cal E}}               % vielbein det in lorentzspace
\newcommand{\W}  {{\cal W}}               % general weights
\newcommand{\OA}  {\Omega}                % chern simons multiplet
\newcommand{\Si}  {\Sigma}                % komplex kaehler multiplet
\newcommand{\GA}{\Gamma}                  % dual G function
\newcommand{\T}  {\theta}

\newcommand{\A}  {\alpha}
\newcommand{\B}  {\beta}
\newcommand{\E}  {\varepsilon}

\newcommand{\Om}  {\omega}

\newcommand{\Sm} {\sigma}
\newcommand{\C} {\chi}

\newcommand{\PAR}[1] {\partial_{#1}}         %         Partial Derivative

\begin{document}
%------------------------------------------------------------------------------

%-------------------------TITLEPAGE-------------------------------------
\thispagestyle{empty}
\begin{titlepage}
\begin{flushright}
%DESY 95--???\\
HUB--EP--95/19 \\
hep-th/9510022 \\
\end{flushright}
\vspace{0.3cm}
\begin{center}
\Large \bf On-Mass-Shell Gaugino Condensation
           \\ in $Z_{N}$ Orbifold Compactifications
\end{center}
\vspace{0.5cm}
\begin{center}
Ingo \ Gaida$^{\hbox{\footnotesize{1,2}}}$ and
Dieter \ L\"ust$^{\hbox{\footnotesize{1}}}$,  \\
{\sl Institut f\"ur Physik, Humboldt--Universit\"at,\\
 Invalidenstrasse 110, D--10115 Berlin, Germany}
\end{center}
\vspace{0.6cm}

\begin{abstract}
\noindent
We discuss non-perturbative aspects
of string effective field theories with $N=1$ supersymmetry
in four dimensions.
By the use of a scalar potential, which is
on-shell invariant under the supersymmetric duality of the dilaton,
we study gaugino condensation in $(2,2)$ symmetric
$Z_{N}$ orbifold compactifications.
The duality under consideration
relates a two-form antisymmetric tensor to a pseudoscalar.
We show, that our approach is independent of the
superfield-representation of the dilaton and preserves the
$U(1)_{PQ}$ Peccei-Quinn symmetry exactly.
\end{abstract}

\vspace{0.3cm}
\footnotetext[1]{E-MAIL: gaida@qft2.physik.hu-berlin.de,
                         luest@qft1.physik.hu-berlin.de.}
\footnotetext[2]{Supported by Cusanuswerk}
\vfill
\end{titlepage}

%-----------------END OF TITLEPAGE----------------------------------------

\setcounter{page}{1}

%---------------------------------------------------------------
%
%  FILE FOR PAPER  DUALITY CONSISTENT GAUGINO CONDENSATION
%  AND SUPERSYMMETRY BREAKING
%
%   ON-MASS-SHELL CASE
%---------------------------------------------------------------

\resetcounter

\setcounter{section}{1}

We discuss effective quantum field theories (EQFT's) of strings
with local N=1 supersymmetry in four dimensions.
These theories are effective in the sense, that they are
low-energy limits of a given higher dimensional string theory
after dimensional reduction and integrating out all heavy modes.
We restrict ourselves to the case of EQFT's, which are only
of second order in derivatives in the bosonic fields.
In these EQFT's the tree level gauge coupling constant is dynamical
and can be expressed
by the vacuum expectation value of the dilaton
superfield. The dilaton superfield can be represented
by the more familiar chiral superfield $S$ or the linear superfield
$L$: $g_{tree}^{2} = 2 \ \vev{S+ \bar S}^{-1} = \vev{L}$.
Throughout this paper $S+ \bar S$ will be denoted as
the chiral representation of the dilaton (S-representation)
and $L$ as the linear representation (L-representation).

It has been shown
that these two superfield representations of the dilaton are
connected via a supersymmetric legendre transformation
called {\em supersymmetric duality}
[\ref{linear}, \ref{Grimm1}, \ref{Chern_Simons},
\ref{Cardoso}, \ref{derendinger}, \ref{derque}].
This duality transformation destroys the holomorphic
structure of the tree level gauge coupling constant.
We will focus on the supersymmetric duality of the dilaton as
an on-shell duality. In the simplest form it
relates the two-form antisymmetric tensor $b_{mn}(x)$
to a pseudoscalar, the so called axion $a(x)$,
via the following algebraic
equation:

%-----------------------------------------
\begin{eqnarray}
\label{comp_duality_trafo}
   \partial_{m} \ a(x)  &=& - \ \E_{mnpq} \  \partial^{n} \ b^{pq}(x)
\end{eqnarray}
%-----------------------------------------

The two-form antisymmetric tensor
is a physical degree of
freedom and plays an important role in the
four dimensional formulation of the Green-Schwarz
anomaly cancellation mechanism
[\ref{Cardoso},\ref{derendinger}].

Since one integrates out only the massive states to go to
the effective theory, the universal degrees of freedom,
namely the graviton, the antisymmetric tensor and the
dilaton, appear in the low energy supergravity action
of four-dimensional, $N=1$ supersymmetric heterotic strings.
Due to phenomenological reasons $N=1$ supersymmetry must be broken and
gaugino condensation
provides a promising mechanism
for spontaneous supersymmetry breaking [\ref{gaugino}].
At the level of an string effective supergravity action this has been
studied extensively
in the $S$-representation of global and local supersymmetry
[\ref{filq},\ref{fmtv},\ref{dual}].

Because the $L$-representation contains a two-form
antisymmetric tensor this formulation seems to be the more natural
EQFT of strings.
It has been shown, that this EQFT most directly reproduces results calculated
in the underlying string theory
[\ref{mayr}, \ref{antoniadis_1}].
Thus, it is an important task to formulate supersymmetry breaking
by gaugino condensation in the linear
representation of the dilaton.
This issue was recently discussed by us
[\ref{gaida}], for some other interesting considerations see also
[\ref{derque}, \ref{binetruy_gaillard}].

In this context we have to face a puzzle: Gaugino condensation
is known to be a non-perturbative effect producing an effective
scalar potential $V \sim e^{-1/g^{2}}$. This can be achieved
in the $S$-representation at tree level by a non-perturbative superpotential
$\Om_{np} \sim e^{-S}$, because $S$ is a chiral superfield.
This is - first of all - impossibel in the $L$-representation,
because the linear multiplet is not chiral and can therefore not
enter the superpotential. On the other hand the duality
(\ref{comp_duality_trafo}) is independent of the dilaton
at component level and contains derivatives. How can then the
$e^{-1/g_{tree}^{2}}$ dependence of the effective scalar potential
be spoilt by the duality transformation?

We will show in the following that the formation of gaugino condensates
can be consistently formulated in both representations of the dilaton.
However, our approach only works on-shell. That is to say, the off-shell
structure of the discussion given here is
still an open problem in many aspects.
One important result of our approach is, that
the well-known $U(1)_{PQ}$ Peccei-Quinn
symmetry is exactly preserved:

%-----------------------------------------
\begin{eqnarray}
\label{pq_symmetry_1}
 b_{mn}(x) \ &\rightarrow& \ b_{mn}(x)
                       \ +  \  \partial_{m} \ b_{n} (x)
                       \ -  \  \partial_{n} \ b_{m} (x)
\\
\label{pq_symmetry_2}
a(x) \ &\rightarrow& \ a(x) \ + \ \Theta  \hspace{2cm} , \ \ \Theta \in {\bf R}
\\
\label{pq_symmetry_3}
 b_{mn}(x) \ &\rightarrow& \ b_{mn}(x)
                       \ +  \ c_{mn} \hspace{1,2cm} , \ \  c_{mn} \in {\bf R}
\end{eqnarray}
%-----------------------------------------

Note that (\ref{pq_symmetry_1}) is known to be the gauge symmetry
of a two-form antisymmetric tensor, whereas (\ref{pq_symmetry_2})
and (\ref{pq_symmetry_3}) are global shift symmetries of
the axion and the antisymmetric tensor respectively.
The on-shell duality transformation (\ref{comp_duality_trafo}) is invariant
under the $U(1)_{PQ}$ symmetry.

The paper is organized as follows:
After a short introduction of
various $N=1$ off-shell multiplets
we discuss the supersymmetric duality
of the dilaton.
These results are already well-known.
Then we derive a duality-invariant scalar potential
and study gaugino
condensation in $(2,2)$ symmetric $Z_{N}$ orbifold
compactifications.
We present a duality-invariant discussion of
gaugino condensation for several gauge groups\footnote{ We
use the convention $\kappa^{2} = 8 \pi / M_{pl}^{2} = 1$.
Moreover we use the usual
superspace notations $\int \equiv \int d^{4}\theta$ and
$X_{|} \equiv X_{| \underline\theta = 0}$ with
$\int d^{4}\theta = - \frac{1}{4} \int d^{2}\theta (\bar \D^{2} - 8R)$.
We will refer to $ \int d^{4}\theta$ as a D-density
and to $ \int d^{2}\theta$ as a F-density.
}.

\vspace{1cm}

%---------------------------------------------------------------------
%---------------------------------------------------------------------

A general supersymmetric multiplet $Z$ has the following structure
at component level:
$Z \sim (\mbox{Bosons} \ | \ \mbox{Fermions} \ || \ \mbox{Auxiliary-Fields} )$.
One of the basic objects in any superspace formulation
[\ref{Grimm1},\ref{wess_and_bagger},\ref{Grimm_2}] are
chiral $4_{B} + 4_{F}$ multiplets
$ \Si \sim  (  A \ | \  \C_{\A} \ || \  F ) $,
because their
lowest components are scalar fields parametrizing
a K\"ahler manifold [\ref{zumino}].
These chiral superfields obey the constraint
$\bar\D^{\dot\A} \ \Si = 0$
and are defined at component level as

%-----------------------------------------
\begin{eqnarray}
\Si_{|} &=& A(x)
\hspace{1cm}
\D_{\A} \Si_{|} = \sqrt{2} \ \C_{\A}(x)
\hspace{1cm}
\D^{2} \Si_{|} = - 4 \ F(x).
\end{eqnarray}
%-----------------------------------------

We will denote in the following
all possible chiral superfields
by $\Si$ except the dilaton in the chiral representation.
The  $4_{B} + 4_{F} $ Yang Mills multiplet
$W_{\A}^{ \ (r)} \sim  ( a_{m}^{ \ (r)} \ | \  \lambda_{\A}^{ \ (r)}  \ || \
D^{(r)}  )$ with the index
$r$ belonging to the internal gauge group $G_{(r)}$
is defined as

%-----------------------------------------
\begin{eqnarray}
W_{\A |}^{ \ (r)} &=& -i \ \lambda_{\A}^{ \ (r)}(x)
\hspace{1cm}
\D_{\B} W_{\A |}^{ \ (r)}  = - \E_{\B\A} D^{(r)}(x)
  -i \ (\Sm^{mn}\E)_{\B\A} \ f_{mn}^{ \ \ (r)}(x).
\end{eqnarray}
%-----------------------------------------
The  Yang-Mills prepotential $V$ satisfies
$W_{\A} = - \frac{1}{4} \ (\bar\D^{2} - 8 R) \
             e^{-2V} \ \D_{\A} \ e^{2V} $.
By constructing the chiral density
$\EPS = e + i e \T \Sm^{a} \bar\psi_{a} - e \T^{2} (\bar M +
        \bar\psi_{a}\bar\Sm^{ab}\bar\psi_{b})$
one finds the $12_{B} + 12_{F} $ minimal multiplet
for the supergravity sector [\ref{minimal_multiplet}],
which we denote by the supercurvature
$R \sim  ( e_{m}^{ \ a}  \ | \  \psi_{m}^{ \ \A}   \ || M, \ b_{a}  )$,
namely the graviton, the gravitino and
two auxiliary fields [\ref{wess_and_bagger}].
The reducibel $8_{B} + 8_{F} $ linear multiplet

%-----------------------------------------
\begin{eqnarray}
L &\sim& ( C, b_{mn}, a_{m}^{ \ (r)} \ | \
           \varphi_{\A}, \lambda_{\A}^{ \ (r)}  \ || \
           D^{(r)}  )
\end{eqnarray}
%-----------------------------------------

is the difference of the $4_{B} + 4_{F} $ Chern-Simons
superfield
$ \OA \sim (  a_{m}^{ \ (r)} \ | \ \lambda_{\A}^{ \ (r)}  \ || \ D^{(r)}  ) $
and the $4_{B} + 4_{F} $ real linear multiplet
$ l \sim ( C, b_{mn} \ | \ \varphi_{\A} \ || \ -)$.
It satisfies the following two constraints:

%-----------------------------------------
\begin{eqnarray}
\label{linear_multiplet_constraint}
 (\bar\D^{2} - 8 R) \ L = -2 \  k_{(r)} \ \ W^{\A (r)} W_{\A (r)}
\hspace{2cm}
L  = l - \  k_{(r)} \ \OA^{(r)} = L^{+}
\end{eqnarray}
%-----------------------------------------

The parameter $k_{(r)}$ denotes the normalization of the
gauge group generators $tr \ T_{(r)} T_{(s)} = k_{(r)} \ \delta_{(r)(s)}$
and is in the context of string theory the level of the
Kac-Moody current algebra of $G_{(r)}$.
The linear mutliplet $L$ is the $N=1$ limit
of the $N=2$ vector-tensor multiplet [\ref{vector_tensor}].
Both multiplets have the same field content,
but the $N=2$ vector-tensor multiplet
is irreducibel because of the extended supersymmetry.
The reducibel parts of the linear superfield, namely the real
linear superfield $l$ and the Chern-Simons superfield $\OA$, satisfy
$( \bar\D^{2} - 8 R ) \ l =  0$ and
$(\bar\D^{2} - 8 R) \ \OA =  2 \ \mbox{tr} \ W^{\A} W_{\A}$
respectively.
This can be used to
write a local F-density into a local D-density up to total
derivatives:
Consider the following Yang Mills action
with an arbitrary chiral function $F(\Si)$
%-----------------------------------------
\begin{eqnarray}
\label{test_action}
  \int d^{2} \T  \  F(\Si)  \ W^{\A} W_{\A}
                       +  \mbox{total derivatives}   +   h.c.
  =  -2  \int d^{4} \T
                   \left \{
                      F(\Si)   +    \bar F( \bar\Si)
                   \right \}  \OA
\end{eqnarray}
%-----------------------------------------

The RHS of (\ref{test_action}) is {\em exactly}
invariant under the following shift
$ F(\Si) \rightarrow  F(\Si) \ + \ i \ \Theta $.
On the LHS of (\ref{test_action}) this holds only if one takes
the boundary terms into account. The LHS of (\ref{test_action})
{\em without} boundary terms
contains at component level the CP-odd term  $f \tilde f$
coupled to the axion.
In perturbation theory these boundary
terms can be ignored, but non-perturbatively this is not
obvious. This has to be taken into account
in the discussion of gaugino condensation.
Furthermore we want to mention, that the
real Yang-Mills Chern-Simons superfield $\OA$ and the
chiral Yang-Mills superfield $ W^{\A} $ have the same field content.
That is why the superfield representation of
the component fields
$(  a_{m}^{ \ (r)} \ | \ \lambda_{\A}^{ \ (r)}  \ || \ D^{(r)}  ) $
is not unique.

The linear multiplet $L$ contains a real scalar $C$, which is called
dilaton in this framework,
its supersymmetric partner, the dilatino $\varphi_{\A}$,
a two-form antisymmetric tensor $b_{mn}$ and the
Yang-Mills Chern-Simons three-form
$\Om_{3Ynml} = - \mbox{tr} (a_{[l}\partial_{m}a_{n]}
               - \frac{2i}{3} \ a_{[l} a_{m} a_{n]} )$
:

%-----------------------------------------
\begin{eqnarray}
\mbox{ln} L_{|} &=& C(x)
\nonumber\\
\nonumber\\
\D_{\A} \ \mbox{ln} L_{|}  &=& \varphi_{\A}(x)
\nonumber\\
\nonumber\\
\com{\D_{\A}}{\bar\D_{\dot\A}} L_{|}
&=& - \frac{4}{3}  e^{C} b_{\A \dot\A}
    + 4 \  k_{(r)} \  \lambda^{(r)}_{\A} \bar\lambda_{\dot\A (r)}
   + \Sm_{k \A \dot\A}
   \left \{
        \E^{klmn} (\partial_{n} b_{ml}
        - \frac{1}{3} \  k_{(r)} \ \Om_{3Ynml}^{(r)}
   \right .
\nonumber\\ & &
   \left .
        + i \ e^{C} \psi_{n} \Sm_{m} \bar\psi_{l})
        +2i \ e^{C} ( \psi_{m} \Sm^{mk} \varphi -
                      \bar\psi_{m} \bar\Sm^{mk} \bar\varphi)
   \right \} (x)
\end{eqnarray}
%-----------------------------------------

Note that $L$ is invariant under the $U(1)_{PQ}$ symmetry.
The duality transformed linear multiplet will
be denoted as $S_{R} = S + \bar S$, where $S$ is a chiral
multiplet. We define

%-----------------------------------------
\begin{eqnarray}
S_{|} &=& (e^{-C} \ + \ i \ a )(x)
\hspace{1cm}
\D_{\A} S_{|} = \sqrt{2} \ \rho_{\A}(x)
\hspace{1cm}
\D^{2}  S_{|} = - 4 \ f(x)
\end{eqnarray}
%-----------------------------------------

In general the gauge group G has a product structure $G = \prod_{(r)} G_{(r)}$.
Nevertheless for  $G_{(r)}$ the gauge coupling is defined
at tree level as $g_{(r)}^{-2} =  k_{(r)} \ \vev{e^{-C}}$.
So these physical parameters of the different
gauge groups $G_{(r)}$ are related to each
other at tree level [\ref{ginsparg},\ref{GHMR}].
The combination $S + \bar S$ is invariant under
the $U(1)_{PQ}$ symmetry
because it depends on the axion only due to derivative terms.
On shell
the supersymmetric
duality of the dilaton
transforms the $2_{B} + 2_{F}$ real linear multiplet
 $ l \sim (  C, b_{mn} \ | \  \varphi_{\A} \ || \ -  ) $
to the  $2_{B} + 2_{F}$ multiplet
$ S_{R} \sim (  C, \partial_{m} a \ | \  \rho_{\A} \ || \ -  )$
without any difficulties. However, the off-shell structure of this duality
relating two inequivalent off-shell theories to each other is still
an open problem in many aspects, although the duality transformation
can be performed off-shell.

All the multiplets we have introduced so far, except the
linear multiplet, are irreducible.
In the context of string compactifications
it is interesting to build reducibel multiplets out of
them and study their relationship to EQFT's with extended
supersymmetry.
We have already mentioned, that the linear multiplet is
associated to the $N=2$ vector-tensor multiplet.
Furthermore the minimal multiplet and the real linear multiplet
form the $16_{B} + 16_{F}$ multiplet
 $ (R + l) \sim  ( e_{m}^{ \ a}, C, b_{mn}   \ | \
            \psi_{m}^{ \ \A}, \varphi_{\A}    \ || \
           M, \ b_{a} ) $,
which was shown to be the $N = 1$ limit of a $N = 4$
EQFT [\ref{nicolai_1}].
In the end the minimal multiplet and the linear multiplet can
combine to the $20_{B} + 20_{F}$ multiplet

%-----------------------------------------
\begin{eqnarray}
\label{enlarged_multiplet}
 (R + L) &\sim& ( e_{m}^{ \ a},  a_{m}^{ \ (r)}, C, b_{mn} \ | \
            \psi_{m}^{ \ \A},  \lambda_{\A}^{ \ (r)}, \varphi_{\A} \ || \
           M, \ b_{a}, D^{(r)}  ).
\end{eqnarray}
%-----------------------------------------

We will couple the irreducible parts of this  $20_{B} + 20_{F}$ multiplet
to matter via the K\"ahler potential
$ K(\Si, \bar\Si)$. The K\"ahler potential is as usual a real function
depending on the supersymmetric matter multiplets $\Si$ of
the underlying EQFT.

In the $U_{K}(1)$-superspace formulation
of $N = 1$ supergravity
[\ref{Grimm1},\ref{Grimm_2}]
the action contains three parts

%-------------------------U(1) LAGRANGIAN------------------
\begin{eqnarray}
\label{U1_lagrangian_general}
 \lagrange &=&  \lagrange_{matter} +
                \lagrange_{pot} +
                \lagrange_{YM}
\end{eqnarray}
%-----------------------U(1) LAGRANGIAN----------------------

with the three basic functions, namely the K\"ahler potential $K$,
the superpotential $\Om$ and the gauge kinetic function
$f_{(r)(s)}$:

%-------------------------U(1) LAGRANGIAN------------------
\begin{eqnarray}
\label{U1_lagrangian}
 \lagrange_{matter}&=&  m_{i}  \int \ E[K_{i}]
\nonumber\\
 \lagrange_{pot}   &=&\frac{1}{2} \int \frac{E}{R} \ e^{K/2} \ \Om(\Si)
                      + h.c.
\nonumber\\
 \lagrange_{YM}    &=& \frac{1}{2} \int \frac{E}{R}
                       \ W^{(r)} \ f_{(r)(s)} \ W^{(s)}
                       + h.c.
\end{eqnarray}
%-----------------------U(1) LAGRANGIAN----------------------

The parameter $m_{i}$ depends on the representation of the dilaton:
It is useful in the following
to introduce the parameter $n = 1/4$.
Then  $m_{i}$ is given as $ m_{linear} = 4n - 3 $
and $m_{chiral} = - 3 $.

Moreover we will use the following variations in
$U_{K}(1)$-superspace for a general superfield $Z$ :

%-----------------------------
\begin{eqnarray}
 \delta_{U} Z  &=& -\frac{\W_{K}(Z)}{2m} \  \frac{\partial K}{\partial U} \
                   \delta U \ Z,
\end{eqnarray}
%-----------------------

The K\"ahler weights
are given as $\W_{K}(E,L,l,\Omega,S,\Si, Y^{3}) = (-2,2,2,2,0,0,2)$, whereas
the superfield $Y^{3} = W^{\A} W_{\A}$ plays an important role
in the context of gaugino condensation, because
its lowest component is given by
gaugino bilinears. In our superspace formulation the
following identity holds:
$ \delta_{Y^{3}} \left ( Y^{3} \ e^{-K/2} \ \right ) = \delta Y^{3} e^{-K/2}$.
And by the use of (\ref{linear_multiplet_constraint}) we find

%-----------------------------
\begin{eqnarray}
\label{key_variation}
 \delta_{Y^{3}} \int d^{4}\theta \ E[K] \ f(\OA)
 = -\frac{1}{2} \int d^{2}\theta \ E[K] \ \delta Y^{3}
    \left \{
        \frac{\partial f(\OA)}{ \partial \OA }
      - \frac{\W_{K}(E)}{2m} \frac{\partial K}{ \partial \OA } f(\OA)
    \right \}.
\end{eqnarray}
%-----------------------

It turns out, that this is a very helpful identity in
the discussion of gaugino condensation.
%----------------------------------------------------------------

\vspace{1cm}

%---------------------------------------------------------------
Before we will derive a general, duality-invariant
scalar potential, we want to have a first look at the duality
(\ref{comp_duality_trafo})
for an ordinary bosonic QFT in the presence of Chern-Simons
forms:
We start with an unconstrained lagrangian

%-------------------------LAGRANGIAN------------------
\begin{eqnarray}
 \lagrange_{u}&=&  \frac{1}{2} \ H^{m} H_{m}
                   \ + \
                   \left (
                      H^{m} \ - \ k_{(r)} \OA^{(r) m}
                   \right ) \partial_{m} a
\end{eqnarray}
%-----------------------LAGRANGIAN----------------------

with $ \OA^{(r) m} = \frac{1}{3}  \E^{mnpq}  \Om_{3Ynpq}$.
Variation with respect to $a(x)$ yields
$ \partial_{m} ( H^{m} -  k_{(r)} \OA^{(r) m} ) =  0$
with the general solution
$H^{m} = \E^{mnpq}   \partial_{n}  b_{pq} +  k_{(r)} \OA^{(r) m}$.
This leads to the following action:

%-------------------------LAGRANGIAN------------------
\begin{eqnarray}
\label{a_action_1}
 \lagrange ( b_{pq}, a ) &=&  \frac{1}{2} \ H^{m} H_{m}
                     +  \partial_{m}
        \left (
                a \  \E^{mnpq} \  \partial_{n} \ b_{pq}
        \right )
\end{eqnarray}
%-----------------------LAGRANGIAN----------------------

Note that (\ref{a_action_1}) is an action of the
antisymmetric tensor only, if one omits the boundary term.
Furthermore $\lagrange (b_{pq})$
is invariant under the $U(1)_{PQ}$ symmetry.
Variation of $\lagrange_{u}$ with respect to $H_{m}(x)$
yields
$ H_{m} =  - \partial_{m} a $.
This leads to a lagrangian, which contains a pseudoscalar
instead of the antisymmetric tensor:

%-------------------------LAGRANGIAN------------------
\begin{eqnarray}
 \lagrange ( \partial_{m} a  ) &=&
%      - \frac{1}{2} \  \partial_{m} a  \ \partial^{m} a
%       - \ k_{(r)} \OA^{(r) m} \ \partial_{m} a
%\nonumber\\ &=&
        - \frac{1}{2} \  \partial_{m} a  \ \partial^{m} a
        -  \ a  \ k_{(r)} \ f_{mn}^{(r)} \ \tilde f^{mn}_{(r)}
        - \ \partial_{m}  \left ( a \ k_{(r)} \OA^{(r) m} \right )
\end{eqnarray}
%-----------------------LAGRANGIAN----------------------

Here we used the usual definition
$ \tilde f_{mn}^{ \ \ (r)} = \frac{1}{2} \E_{mnpq} f^{pq (r)} $
and the well-known identity
$ \partial_{m}  \OA^{(r) m} = - f_{mn}^{(r)}  \tilde f^{mn}_{(r)} $.
Again the resulting lagrangian is invariant under the
$U(1)_{PQ}$ symmetry.
If we consider a constrained lagrangian $\lagrange_{c}$, which
is $\lagrange_{u}$ satisfying
the on-shell constraint:
$ H^{m}  =   \E^{mnpq} \  \partial_{n} \ b_{pq}
              \ + \ k_{(r)} \OA^{(r) m}
         =   - \ \partial^{m} a $,
then we can relate the two models to each other directly -
one containing an antisymmetric tensor and the other
a pseudoscalar.

So far we have only summarized well-known results
that are useful in the following.
We want to discuss the duality now in the framework
of $N=1$ superspace:
The tree-level K\"ahler potential under discussion
in the linear representation of the dilaton will be
$\tilde K = 4n \ \mbox{ln} (L/2) + K(\Si, \bar \Si)$.
It already includes the tree level gauge kinetic function,
but therefore it is not a well-defined K\"ahler potential
in the sense, that it does not fulfill the K\"ahler condition
[\ref{gaida}].
Furthermore at one loop the Wilsonian gauge coupling function
is determined by a holomorphic function of the chiral fields $\Si$:
$f_{(r)(s)}^{[1]} =  \delta_{(r)(s)} \ f^{[1]}(\Si) $.
We assume here, that in perturbation theory the Wilsonian gauge coupling
function is beyond tree level independent of the dilaton [\ref{DKL}].
The lagrangian considered here is evaluated at component level
in [\ref{Grimm1}].
The part of the lagrangian containing the auxiliary fields
reads

%-----------------------------
\begin{eqnarray}
 \lagrange_{aux}/e = \frac{4n-3}{9} M \bar M
                     + G_{i \bar j} F^{i} {\bar F}^{\bar j}
                     + e^{\GA/2} \left (
                                  \frac{4n-3}{3}(M+\bar M)
                                  +  F^{i} G_{i}
                                  + \bar F^{\bar j} \bar G_{\bar j}
                                 \right ),
\end{eqnarray}
%-----------------------

where G is the well-defined dual G-function:
$ G = K(\Si, \bar \Si) + \mbox{ln} |\Om|^{2}$.
So the scalar potential from this part after elimination of the
auxiliary fields is

%-----------------------------------
\begin{eqnarray}
\label{linear_potential_1}
 V_{1}  &=& e^{\GA} \ (G_{i} \ G^{i \bar j} G_{\bar j}  + 4n - 3 )
\end{eqnarray}
%-----------------------------------

The second part of the potential is the sum of all monomials
coupling the dilaton $C$ only to tr $\lambda^2$:

%-----------------------------
\begin{eqnarray}
 V_{2} &=&
%    -2n \ e^{-C} \ e^{\GA/2}
%    (\mbox{tr}\lambda^2 + \mbox{tr}\bar \lambda^2 )
%    -n \ e^{-2C} \ \mbox{tr}\lambda^2 \  \mbox{tr}\bar \lambda^2
%\nonumber\\
% &=&
%    - (n \ \mbox{tr}\lambda^2 + 2n \ e^{C} e^{\GA/2} )
%      \frac{ e^{-2C} }{n}
%      (n \ \mbox{tr}\bar \lambda^2  + 2n \ e^{C} e^{\GA/2} )
%    + 4n \ e^{\GA}
%\nonumber\\
% &=&
      (n \ \mbox{tr}\lambda^2 -  e^{\GA/2} \GA_{L} )
      \ \GA^{LL} \
      (n \ \mbox{tr}\bar \lambda^2  -  e^{\GA/2} \GA_{L} )
      - 4n \ e^{\GA}
\end{eqnarray}
%-----------------------

Derivatives with respect to the linear multiplet
are defined as $\partial_{L} = \frac{\partial}{\partial(2/L)}$.
The whole potential can be rewritten as

\begin{eqnarray}
\label{linear_potential_1}
 V  &=& \left(
            n \ \delta_{iL} \ tr \ \lambda^{\A} \lambda_{\A}
            - e^{\GA/2} \GA_{i}
        \right)
        \ \GA^{i \bar j}
        \left(  n \ \delta_{\bar j L} \
                tr \ \bar\lambda_{\dot\A} \bar\lambda^{\dot\A}
                -  e^{\GA/2} \GA_{\bar j}
        \right)
         - 3 e^{\GA} .
\end{eqnarray}

The indices $i,\bar j$ include derivatives with respect to
the linear muliplet now. And $\delta_{iL}$ is the Kronecker-delta.
The result reduces to the known potential in the chiral limit $n=0$.

Performing now the duality transformation
our lagrangian
changes to an unconstrained lagrangian $\lagrange_{u}$ by changing
$l$ to $U$,
where $U$ is unconstrained, and by adding a lagrange multiplier
$\lagrange_{lm}$.
The lagrange multiplier contains
the unconstrained field $U$ and the $S+\bar S$ multiplet.

%-----------------------------
\begin{eqnarray}
\lagrange_{u} &=&
    m \ \int \ E \
    \left (
           1 - \frac{2n}{m} \ U \ (S+\bar S)
    \right )
    + \lagrange_{pot}
\end{eqnarray}
%-----------------------

Variation with respect to $S+\bar S$ yields the old theory in the
linear representation of the dilaton.
Variation with respect to $U$ yields the chiral representation of the
dilaton.
Note that the variations depend on the representation of the
dilaton.
In the $U_{K}(1)$-superspace the torsion constraints and
consequently the solution of the Bianchi-identities depend
on the representation
of the dilaton: In the notation of [\ref{Grimm_2}]
the superdeterminant of the vielbein $E$, for instance, gets rescaled
as $ E^{\prime}  = E \ (X \bar X)^{2}$ with
$X = \bar X = e^{K/4m}$.
The rescalings are chosen in such a way,
that the whole theory is automatically Einstein normalized.
\footnote{In Lorentz-superspace
the Weyl-rescaling of the graviton depends on the representation
of the dilaton [\ref{gaida}].}
Thus, we find with $\tilde K(U) = 4n \ \mbox{ln} (U - k_{(r)} \Omega^{(r)} )$
the {\em duality relation}
$ S + \bar S  = 2/L$.
Inserting this relation
one ends up with the action in the chiral representation
of the dilaton:

%-----------------------------
\begin{eqnarray}
\label{dual_action}
 \lagrange &=&
 - 3 \ \int E[K]   \left (
                     1 + \frac{2n}{3} k_{(r)} \Omega^{(r)} \ ( S + \bar S)
                   \right )
\nonumber\\ & &
 + \left\{
      \frac{1}{2} \int  \frac{E}{R} \ e^{K/2} \ \Om(\Si)
    + \frac{1}{2} \int  \frac{E}{R} \ W^{(r)} \
      f_{(r)(s)}^{ \ \ [1]} \ W^{(s)} \  + h.c.
   \right\}
\end{eqnarray}
%-----------------------

This action is manifestly invariant
under the $U(1)_{PQ}$ symmetry
of the dilaton in the chiral representation. It depends only on
$\partial_{m} a (x)$, because the theory in the
$L$-representation only knows about the field strength
of the antisymmetric tensor.
One open problem is encoded in the fact, that off-shell we have in the
$S$-representation the corresponding
auxiliary field $f$, which is absent in the linear
representation. But since we are discussing the supersymmetric duality
only on-shell, we have to eliminate the auxiliary fields via their
equations of motion.
It is quite interesting, that the remaining scalar potential
is the same than the one in the $L$-picture as we will show now:
The K\"ahler potential and the
gauge coupling function are given in the $S$-representation
by $K  = -4n \ \mbox{ln} (S + \bar S) + K(\Si, \bar \Si)$
and
$f_{(r)(s)} =    n \ \delta_{(r)(s)} \ \left( f^{[0]} + f^{[1]} \right)$
respectively.
The tree-level gauge coupling function
$ f^{[0]} = (S + \bar S) \ k_{(r)} = 2 \ k_{(r)}/L$
is part of a D-density.
This point of view was already
very successful in the discussion of non-holomorphic
field-dependent contributions to the gauge coupling function
[\ref{derendinger}].
After performing the usual K\"ahler transformation
to go to the G-function the auxiliary part of the lagrangian is given by

%-----------------------------
\begin{eqnarray}
 \lagrange_{aux}/e &=& - \frac{1}{3} M \bar M
                     + G_{i \bar j} F^{i} {\bar F}^{\bar j}
                   - n \ F^{i} \ \delta_{iS} \ tr \ \lambda^{\A} \lambda_{\A}
                   - n \ \bar F^{\bar j} \ \delta_{\bar j \bar S}
                           \ tr \ \bar\lambda_{\dot\A} \bar\lambda^{\dot\A}
\nonumber\\ & &
                   + e^{G/2}
                \left (
                       F^{i} G_{i}
                       + \bar F^{\bar j} \bar G_{\bar j}
                       - M - \bar M
                \right ),
\end{eqnarray}
%-----------------------

with $ F^{i} \ \delta_{iS} = f $.
Eliminating the auxiliary fields via their equation of motion
leads to the scalar potential:

\begin{eqnarray}
\label{chiral_potential_1}
 V  &=& \left(
            n \ \delta_{iS} \ tr \ \lambda^{\A} \lambda_{\A}
            - e^{G/2} G_{i}
        \right)
        \ G^{i \bar j}
        \left(  n \ \delta_{\bar j \bar S} \
                tr \ \bar\lambda_{\dot\A} \bar\lambda^{\dot\A}
                -  e^{G/2} G_{\bar j}
        \right)
         - 3 e^{G}
\end{eqnarray}

This potential includes the usual scalar potential of ordinary matter fields
of [\ref{sugra_1}] and is precisely (\ref{linear_potential_1}) -
the scalar potential derived in the
linear representation of the dilaton.
%------------------------------------------------------------------

\vspace{1cm}

%-----------------------------------------------------------------
Now we can use the duality-invariant scalar
potential to study gaugino condensation in
the two superfield-representations of the dilaton.
We will restrict ourselves first of all
to the case of one gauge group and take
$k_{(r)} = 1$ for simplicity.
We start in the linear representation of the dilaton:
The superpotential consists of two parts:
The first part is the so called quantum  part
and has its origin in chiral and conformal
anomalies [\ref{taylor1}].
The second part represents the one-loop threshold corrections
to the gauge coupling function.
Note that the superpotential is explicitly dilaton-free, because
it is defined to be a chiral function. However
it depends implicitly on the dilaton through the K\"ahler potential,
which can enter the superpotential\footnote{This is also
the case in the superconformal approach, where the compensators
are functions of the K\"ahler potential.}.
The two parts combine to the following effective superpotential

%-----------EFFECTIVE SUPERPOTENTIAL---------------------------
\begin{eqnarray}
\label{effective_superpotential}
  \Om(\Si) &=&  \frac{1}{\B} \ Y^{3} e^{- \tilde K/2} \
             \mbox{ln}
             \left \{
                      c^{6}  \ e^{ \B n f^{[1]} } \ Y^{3} \ e^{- \tilde K/2}
             \right \}
\end{eqnarray}
%-----------END OF EFFECTIVE SUPERPOTENTIAL--------------------
with $ \B = -24 \pi^{2}/b \ n $, where $b$ denotes
the $N=1$ $\B$-function coefficient.
By the use of (\ref{key_variation}) we find the equations of motion
for $Y^{3}$.
Because the action contains the Yang-Mills Chern-Simons superfield
$\OA$ and its chiral projection $Y^{3}$ at the same
time, the equation of
motion splits into a non-holomorphic and a holomorphic part.
This property can be used to introduce non-holomorphic terms
into the superpotential. After an appropriate K\"ahler transformation
with $\tilde K \rightarrow \Gamma = \tilde K + \mbox{ln} |\Om|^{2}$
we have

%---------------------------------------------------------------------
\begin{eqnarray}
\label{gaugino_constraint_01}
  \lambda^{\A}\lambda_{\A}  &=&  e^{\Gamma/2} \ \frac{1}{1/ \B + n f^{[0]} }.
\end{eqnarray}
%---------------------------------------------------------------------

So we find immediately, that in the weak coupling limit
$f^{[0]} \rightarrow \infty$ the vacuum expectation value of
the gaugino bilinears vanishes and consequently gaugino
condensation does not take place.
The scalar potential is given by (\ref{linear_potential_1})
and by the use of the equations of motion for the gaugino bilinears we can
integrate them out. At this point we want to show, that the
results given here are invariant under the duality transformation:
Performing the duality transformation the action is given by
(\ref{dual_action}).
Again the tree-level gauge coupling function
is part of a D-density.
Analogous to the linear representation of the dilaton
the effective superpotential is given by (\ref{effective_superpotential}).
Using (\ref{key_variation}) we find the equation of motion
for $Y^{3}$ in the $S$-representation. This yields
again (\ref{gaugino_constraint_01}) with $G$ instead of $\Gamma$,
of course.
Again it is now straightforward to integrate out the
gaugino bilinears using (\ref{chiral_potential_1}).
In the end the procedure of integrating out the gaugino bilinears
via their equations of motion is
not affected by the duality transformation.
%---------------------------------------------------------

\vspace{1cm}

%--------------------------------------------------------
As a concrete example we turn now to the discussion of
orbifold models describing the
compactification of the heterotic string from ten dimensions down to
four [\ref{orbifold_ref}].
We will focus on (2,2) symmetric $Z_{N}$ orbifold
compactifications without Wilson lines. In these models
occurs the generic gauge group
$E_{8} \ \bigotimes \ E_{6} \ \bigotimes \ H$
with $H = \{SU(3), SU(2) \times U(1), U(1)^{2} \}$.
Orbifold compactifications posses various continuous parameters,
called moduli, corresponding to marginal deformations of
the underlying conformal field theory. These moduli enter the
EQFT and they take their values in a manifold $\cal M$ called
moduli space. For models with $N = 1$ space-time supersymmetry
$\cal M$ is, locally, a K\"ahlerian manifold. Thus, the corresponding
K\"ahler potential describes the coupling of the moduli to
$N = 1$ supergravity in the EQFT under consideration.
The EQFT must respect
target space modular symmetries (for a review see [\ref{gpr}]) induced
by the target space duality group. For orbifold compactifications
the target space duality group is often given by the modular group
$PSL(2,{\bf Z})$,
acting on one chiral field $T$ as

\begin{eqnarray}
\label{modular_transformation_of_moduli}
 T^{\prime}   &=& \frac{a \ T - i \ b}{ i \ c \ T + d } \hspace{1cm}
       ad - bc     =  1 \hspace{1cm}  a,b,c,d \in {\bf Z}
\end{eqnarray}

where $T$ corresponds to an internal, overall modulus: $T=R^2+iB$.
For simplicity we will discuss (2,2) symmetric $Z_{N}$ orbifolds without
(1,2) moduli.
At the massless level of these orbifolds one finds
(1,1) moduli and matter fields. These fields can be in
general (un-) twisted and (un-) charged.
In the following the three diagonal elements of the untwisted uncharged
(1,1) moduli are denoted by $T^{A}$, whereas all other
charged (uncharged) fields are given by $Q_{ch}^{I} $
($Q_{uch}^{I}$). We will only be interested
in the lowest components of the superfields in order to study
the vacuum structure of our theory. Since in the Wess-Zumino gauge
we have $ V_{| WZ} = 0 $, it is not necessary for us to
investigate the role of the charged and uncharged matter
separately. So we define $Q^{I} = \{ Q_{ch}^{I}, Q_{uch}^{I} \}$
and use this definition in an obvious way.

Now we must specify the K\"ahler potential $\tilde K$
and the
effective superpotential $\Om$.
The K\"ahler potential $\tilde K$ can be generically expanded
in powers of $Q^{I}$:

\begin{eqnarray}
\label{orbifold_kaehlerpotential}
 \tilde K &=& \mbox{ln}(L/2) \ + \ \hat K(T) \ + \ Z_{I \bar J} (T , \bar T)
              \ \bar Q^{\bar J} \ e^{2V} \ Q^{I}
              \ + \ {\cal O}
              \left(
                (  \bar Q  Q)^{2}
              \right )
\end{eqnarray}

We choose the matter metric in the following diagonal form

\begin{eqnarray}
\label{orb_kaehler_2}
  Z_{I \bar J} (T, \bar T) = \delta_{I \bar J} \ Z^{I}  (T, \bar T)
\hspace{2cm}
  Z_{I} (T, \bar T) = \prod_{A} \ ( T^{A} + \bar T^{A} )^{-q_{I}^{ \ A}},
\end{eqnarray}

whereas the metric for the moduli is given by
$\hat K_{i \bar j}$  with
$\hat K =  -  \sum_{A=1}^{3} \ \mbox{ln}(T^{A}+\bar T^{A} )$.
We will use the usual notation
$K^{i \bar j} = K_{i \bar j}^{-1}$ also for the matter metric.
The parameters $q_{I}^{ \ A}$ are the so called modular weights
[\ref{mod_weights}].
The superpotential has the following structure

\begin{eqnarray}
\label{orbifold_superpotential}
  \Om (\Si) &=& P(Q,T) \ + \ \sum_{(r),A}  \Om_{(r)}^{ \ A} (\Si).
\end{eqnarray}

More precisely we have superpotentials of the following form in mind

\begin{eqnarray}
\label{orbifold_superpotential_2}
  P(Q,T) &=&  \frac{1}{3} \ Y_{IJK}(T) \ Q^{I} \ Q^{J} \ Q^{K}
\\
\label{orbifold_superpotential_3}
  \Om_{(r)}^{ \ A} (\Si) &=& \frac{1}{ \B_{(r)}^{A} }
                             \ Y^{3} e^{-\tilde K/2} \
             \mbox{ln}
             \left \{
                      c_{(r)}^{ \ A 6} \ Y^{3} \
                      e^{-\tilde K/2} \ e^{n \B_{(r)}^{A} f^{[1] A}_{(r)}}
             \right \},
\end{eqnarray}

with the definition $\B_{(r)}^{A} = - 24 \pi^{2} / b_{(r)}^{A} n $.
The canonical dimensions of the fields are
$dim (T^{A}, l, \OA, Y, Q^{I}) = (0, 0, 2, 1, 1)$
and the transformation properties under the target space
duality group
(\ref{modular_transformation_of_moduli})
with $F_{A} = \mbox{ln} (icT_{A} + d)$ read

\begin{eqnarray}
\label{orbifold_kaehlerpotential_2}
   T^{A} + \bar T^{\bar A}
&\rightarrow&
   (T^{A} + \bar T^{\bar A}) \ e^{- (F_{A} + \bar F_{A})}
\hspace{1cm}
\mbox{(no sum over $A$)}
\nonumber\\
\nonumber\\
 Z_{I \bar J} (T, \bar T)
&\rightarrow&
    Z_{I \bar J} (T, \bar T) \ e^{q_{I}^{ \ A}(F_{A} + \bar F_{A})}
\hspace{1cm}
\mbox{(sum over $A$)}
\nonumber\\
\nonumber\\
 Q_{I}
&\rightarrow&
    Q_{I} \ e^{- q_{I}^{ \ A} F_{A} }
\hspace{2,8cm}
\mbox{(sum over $A$)}
\end{eqnarray}

Therefore the target space duality transformations act just as
K\"ahler transformations.

The potential is given in general by (\ref{linear_potential_1}).
Note that this is a `closed' formula without
specifying the superpotential:

\begin{eqnarray}
\label{z_n_potential_2}
 G_{i} \ G^{i \bar j} G_{\bar j} &=& \frac{1}{t} \
\left \{
\frac{T_{R}^{ \ A 2}}{t} \ | t \ \PAR{T^{A}} \mbox{ln} \ \Om
                        \ - \ \frac{1}{T_{R}^{ \ A}} |^{2}
\ - \ \frac{1}{t} \ |q_{I}^{ \ A 2} \  Q^{2}  |^{2}
\right .
\nonumber\\ & &
+ \ Z^{I \bar J } \
\left (
q_{I}^{ \ A 2} \ Q^{2}  \  | \PAR{Q^{I}} \mbox{ln} \ \Om |_{I \bar J}^{2}
\ + \ t \ | Z_{I \bar K}  \bar Q^{\bar K} \
        + \ \PAR{Q^{I}} \ \mbox{ln} \ \Om |_{I \bar J}^{2}
\right .
\nonumber\\ & &
- \  | q_{I}^{ \ A} Z_{I \bar K}  \bar Q^{\bar K}
        \ + \ \PAR{Q^{I}} \ \mbox{ln} \ \Om |_{I \bar J}^{2}
\nonumber\\ & &
\left .
\left .
+ \ | q_{I}^{ \ A} Z_{I \bar K}  \bar Q^{\bar K} \
        T_{R}^{ \ A} \PAR{T^{A}} \mbox{ln} \ \Om
        \ + \ \PAR{Q^{I}} \mbox{ln} \ \Om |_{I \bar J}^{2}
\right )
\right \}
\end{eqnarray}

All indices are contracted and the following definitions have
been used:
$ Q^{2} = Z_{I \bar J} (T , \bar T) \bar Q^{\bar J}  Q^{I}$
and
$ t  =  1 - q_{I}^{ \ A 2} \ Q^{2} $.
The equation of motion for the gaugino bilinears yields now

\begin{eqnarray}
\label{gaugino_constraint_orbifold_1}
  tr \  \lambda^{\A}\lambda_{\A}  &=&
    \frac{1}{3} \ e^{\Gamma/2} \ tr \ \sum_{A=1}^{3} \
    \left (
           \frac{1}{ \B_{(r)}^{A} } \ + \ n \ f_{(r)}^{[0]A}
    \right )^{-1}
\end{eqnarray}

with $f_{(r)}^{[0]A} = f_{(r)}^{[0]} /3 = 2k_{(r)}/3L = k_{(r)} S_{R} /3$.
In the weak coupling limit gaugino condensation still disappears.
Using the equation of motion
for the gaugino bilinears leads to non-holomorphic contributions
to the superpotential:

\begin{eqnarray}
\label{truncated_superpotential_orbifold}
 \Om^{A}_{(r)} &=& - e^{-1} \ c^{A \ -6}_{(r)} \
      e^{- n \B^{A}_{(r)} ( f_{(r)}^{[0]A} +  f_{(r)}^{[1]A} ) }
                   \left (
                       \frac{1}{ \B_{(r)}^{A} } \ + \ n \ f_{(r)}^{[0]A}
                   \right )
\end{eqnarray}

It is important to stress, that $f_{(r)}^{[0]A}$ can be expressed
by the use of the equations of motion for $Y^{3}$ in a pure
holomorphic way. This property is directly related to the fact,
that we have to deal with the non-holomorphic Chern-Simons superfield
$\OA$ and its holomorphic projection $Y^{3}$ at the same time.
As a consequence the remaining scalar potential has the
desired non-perturbative structure
$ V \sim e^{-(S + \bar S)} = e^{-2/L} \sim e^{-1/g^{2}_{tree}} $,
if the theory is asymptotical free ($\B_{(r)}^{A} > 0$).
Our approach produces this functional dependence on the
dilaton as an overall factor in (\ref{truncated_superpotential_orbifold}).
So the $U(1)_{PQ}$ symmetry is still unbroken.
The matter fields we have introduced are quantum fields
with vanishing vacuum expectation value. Because
we are interested in the vacuum structure of our
theory, we take the limit $Q^{I} \rightarrow 0$.

Up to now it was not necessary to specify the one-loop
contribution to the gauge-coupling function.
Following [\ref{DKL}] we have

%-------------GAUGE KINETIC FUNCTION-----------------
\begin{eqnarray}
\label{one-loop_gauge_function}
 f^{[1]A}_{(r)}(\Si) &=& - \ \frac{b^{A}_{ \ (r)}}{ 8 \pi^{2}} \
                           \mbox{ln} \ \eta^{2}(T^{A}) ,
\end{eqnarray}
%-----------GAUGE KINETIC FUNCTION-------------------

where $ \eta(T^{A})$ is the well-known Dedekind function
and reflects the one-loop threshold contributions
of momentum and winding states of the underlying string theory.
For simplicity we have assumed, that the coefficient
which appears in the threshold correction is the $N=1$
$\B$-function coefficient. This is the case for a hidden
gauge group of pure Yang-Mills theory like $E_{8}$ in the
absence of Green-Schwarz terms. The inclusion of a Green-Schwarz term
was discussed in the last reference of [\ref{dual}].

Using (\ref{gaugino_constraint_orbifold_1})
the scalar potential is given as

%-------------CHIRAL MODULAR INVARIANT TRUNCATED POTENTIAL---------------------
\begin{eqnarray}
\label{full_linear_potential_0}
 V(C,T^{A}) =
      \frac{ e^{C} }{2} \prod_{A=1}^{3} (T_{R}^{ \ A})^{-1}
      | {\Om}|^{2}
      \left \{
      \sum_{A=1}^{3} T^{ \ A 2}_{R} \
      |\frac{3{\hat G}_{2}(T^{A})}{2 \pi} + \frac{2}{ T_{R}^{ \ A}}|^{2}
              + k(C)  - 3
      \right \}
\end{eqnarray}
%--------END OF CHIRAL MODULAR INVARIANT TRUNCATED POTENTIAL--------------

with the Eisenstein function $\hat{G}_{2}(T) = G_{2}(T) \ - \ 2\pi/T_{R}$
and

%----------------------------------
\begin{eqnarray}
  k(C) &=& | \ tr \ \sum_{A=1}^{3}
           \frac{ \B_{(r)}^{A} }{ 6e^{C} \ + \  \B_{(r)}^{A} \ k_{(r)}}
           + 1 \ |^{2}.
\nonumber\\
\end{eqnarray}
%----------------------------------

We have calculated the effective scalar potential
for factorizable K\"ahlerian moduli spaces ${\cal M}$ of the form
${\cal M} =  {\cal M}^{(1,1)}_{dilaton} \ \bigotimes \  {\cal M}^{\prime}$
with  ${\cal M}^{(1,1)} = SU(1,1)/U(1)$.
It is well-known that supersymmetry can be broken via
effective scalar potentials of the form
(\ref{full_linear_potential_0}) with (\ref{one-loop_gauge_function})
[\ref{filq}]. This property
does not depend on the representation of the dilaton [\ref{gaida}].

\vspace{1cm}

To conclude, we have shown in this paper, that there exists
a consistent on-mass-shell formulation of gaugino condensation
in local $N=1$ string effective field theories in four
dimensions.

We studied the one-loop anomalous contribution to the Wilsonian
gauge coupling, discussed by Dixon, Louis and Kaplunovsky
[\ref{DKL}], and the anomalous contribution to the effective
action of Taylor, Veneziano and Yankielowicz [\ref{taylor1}].
Both can combine to an effective superpotential.
Using this effective superpotential we
studied supersymmetry breaking via gaugino condensation
in (2,2) symmetric $Z_{N}$ orbifolds.
In our approach we have integrated out
the gaugino bilinears via their equations of motion.
By the use of the Yang-Mills Chern-Simons superfield and its chiral
projection the equations of motion split into a
holomorphic and a non-holomorphic part. This result,
which clearly only holds on-shell, leads to the
non-perturbative structure of the effective scalar
potential with the typical $e^{-1/g^{2}_{tree}}$
behaviour. Moreover we have shown that our approach
is independent of the superfield-representation
of the dilaton and preserves
the $U(1)_{PQ}$ symmetry.

Finally, we want to mention that the duality
invariant off-shell formulation
of gaugino condensation is still an open problem in
many aspects.
%-----------------------------------------------------------

\hspace{0.5cm}

{\bf Acknowledgement:} We would like to thank
G. Lopes-Cardoso and C. Preitschopf for the many
conversations and
J.P. Derendinger for helpful discussions.

\hspace{0.5cm}

%---------------------------------------------------------------------
%
%                        BIBLIOGRAPHY
%
%----------------------------------------------------------------------

\section*{References}
\begin{enumerate}
\item
\label{linear}
S. Ferrara, J. Wess and B. Zumino, Phys. Lett. {\bf B51} (1974) 239;
\\
W. Siegel, Phys. Lett. {\bf B85} (1979) 333;
\\
S. Ferrara and M. Villasante, Phys. Lett.{\bf  B186} (1986) 85;
\\
S. Cecotti, S. Ferrara and L. Girardello, Phys. Lett. {\bf B198} (1987) 336;
\\
B. Ovrut, Phys. Lett.{\bf  B205} (1988) 455;
\\
B.A.~Ovrut and C.~Schwiebert, Nucl.Phys. {\bf B321} (1989) 163:
\\
B. Ovrut and S.K. Rama, Nucl. Phys. {\bf B333} (1990) 380;
\\
B. Ovrut and S.K. Rama, Phys. Lett. {\bf B254} (1991) 138;
\\
P. Binetruy,G. Girardi and R. Grimm, Phys. Lett. {\bf B265} (1991) 111.
\item
\label{Grimm1}
P.~Binetruy, G.~Girardi, R.~Grimm and M.~M\"uller,
Phys. Lett. {\bf B195} (1987) 389;
\\
P.~Adamietz,P.~Binetruy, G.~Girardi and R.~Grimm,
Nucl. Phys. {\bf B401} (1993) 275.
\item
\label{Chern_Simons}
S.~Cecotti, S.~Ferrara and M.Villasante, Int.J.Mod.Phys. {\bf A2} (1987)
1839
\\
G.~Girardi and R.~Grimm, Nucl.Phys.  {\bf B292} (1987) 181.
\item
\label{Cardoso}
G.~Lopes-Cardoso and B.~Ovrut, Nucl.Phys.{\bf B369} (1992) 351; \\
G.~Lopes-Cardoso and B.~Ovrut, Nucl.Phys. {\bf B392} (1993) 315
\item
\label{derendinger}
J.P.~Derendinger, S.~Ferrara, C.~Kounnas and F.~Zwirner,
Nucl. Phys. {\bf B372}  (1992) 145.
\item
\label{derque}
J.P.~Derendinger, F.~Quevedo and M.~Quiros,
      Nucl. Phys. {\bf B428} (1994) 282;
\\
C.P.~Burgess, J.P.~Derendinger, F.~Quevedo and M.~Quiros,
       Phys. Lett. {\bf B 348} (1995) 428;
\\
C.P.~Burgess, J.P.~Derendinger, F.~Quevedo and M.~Quiros,
      hep-th 9505171.
\item
\label{gaugino}
H.P. Nilles, Phys. Lett. {\bf B115} (1982) 193;
\\
S. Ferrara, L. Girardello and H.P. Nilles, Phys. Lett. {\bf B125} (1983) 457;
\\
J.P. Derendinger, L.E. Ibanez and H.P. Nilles, Phys. Lett. {\bf 155} (1985) 65;
\\
M. Dine, R. Rohm, N. Seiberg and E. Witten, Phys. Lett. {\bf 156} (1985) 55.
\item
\label{filq}
A.~Font, L.E.~Ib\`a\~nez, D.~L\"ust and F.~Quevedo, Phys. Lett.
{\bf B245} (1990) 401.
\item
\label{fmtv}
S.~Ferrara, N.~Magnoli, T.R.~Taylor and G.~Veneziano, Phys.
Lett. {\bf B245} (1990) 409.
\item
\label{dual}
C.~Kounnas and M.~Porrati, Phys. Lett. {\bf B191} (1987) 91;
\\
N.V. Krasnikov, Phys. Lett. {\bf 193} (1987) 37;
\\
L. Dixon, in Proc. 15th APS D.P.F. Meeting, 1990;
\\
J.A. Casas, Z. Lalak, C. Munoz and G.G. Ross, Nucl. Phys. {\bf 347} (1990) 243;
\\
T.R. Taylor, Phys. Lett. {\bf 252} (1990) 59;
\\
H.-P.~Nilles and M.~Olechowski, Phys. Lett. {\bf B248} (1990) 268;
\\
D. L\"ust and T. Taylor, Phys. Lett. {\bf B253} (1991) 335;
\\
M. Cvetic, A. Font, L.E. Ibanez, D. L\"ust and F. Quevedo, Nucl. Phys.
{\bf B361} (1991) 194;
\\
D. L\"ust and C. Munoz, Phys. Lett. {\bf B279} (1992) 272.
\item
\label{mayr}
P.~Mayr and S.~Stieberger, Nucl. Phys. {\bf B412} (1994) 502.
\item
\label{antoniadis_1}
I.~Antoniadis, E.~ Gava and  K.S.~Narain, Phys. Lett. {\bf B283} (1992) 209.
\item
\label{gaida}
I.~Gaida and D.~L\"ust, Int. J. Mod. Phys. {\bf A10} (1995) 2769.
\item
\label{binetruy_gaillard}
P.~Binetruy, M.K.~Gaillard and T.R.~Taylor,  hep-th 9504143.\\
P.~Binetruy and M.K.~Gaillard,  hep-th 9506207.
\item
\label{wess_and_bagger}
J.~Wess and J.~Bagger, Supersymmetry and Supergravity,
Princeton University.
\item
\label{Grimm_2}
P.~Binetruy, G.~Girardi, R.~Grimm and M.~M\"uller,
Phys.Lett. {\bf B189} (1987) 83;
\\
P.~Binetruy, G.~Girardi and R.~Grimm, LAPP-preprint LAPP-TH-275/90  (1990).
\item
\label{zumino}
B.~Zumino, Phys. Lett. {\bf B87} (1979) 203.
\item
\label{minimal_multiplet}
K.S.~Stelle and P.C.~West, Phys. Lett. {\bf B74} (1978) 330;
\\
S.~Ferrara and P.~Niewenhuizen, Phys. Lett. {\bf B74} (1978) 333.
\item
\label{vector_tensor}
M.~Sohnius, K.S.~Stelle and P.C.~West, Phys. Lett. {\bf B92} (1980) 123;
\\
B.~deWit, V.~Kaplunovsky, J.~Louis and D.~L\"ust,
                         Nucl. Phys. {\bf B451} (1995) 53.
\item
\label{ginsparg}
P.~Ginsparg , Phys. Lett. {\bf B197} (1987) 139.
\item
\label{GHMR}
D.~Gross, J.~Harvey, E.~Martinec and R.~Rohm,
                    Nucl. Phys. {\bf B256} (1985) 253; \\
D.~Gross, J.~Harvey, E.~Martinec and R.~Rohm,
               Nucl. Phys. {\bf B267} (1986) 75.
%\item
%\label{louis_kaplunovsky_2}
%J.~Louis and V.~Kaplunovsky,  Nucl. Phys. {\bf B444} (1995) 191.
\item
\label{nicolai_1}
H.~Nicolai and P.K.~Townsend, Phys. Lett. {\bf B98} (1981) 257.
\item
\label{gpr}
A.~Giveon, M.~Porrati and E.~Rabinovici, Phys. Rep. {\bf 244} (1994) 77.
\item
\label{sugra_1}
E.~Cremmer, S.~Ferrara, L.~Girardello and A.~Van~Proeyen,
Nucl. Phys. {\bf B212} (1983) 413;
\\
E.~Cremmer, B.~Julia, J.~Scherk, S.~Ferrara, L.~Girardello
and P.~Van~Nieuwenhuizen, Nucl. Phys. {\bf B147} (1979) 105.
\item
\label{taylor1}
G.~Veneziano and S.~Yankielowicz, Phys. Lett. {\bf B113} (1982) 231;
\\
T.R.~Taylor, G.~Veneziano and S.~Yankielowicz,
Nucl. Phys. {\bf B218} (1983) 493.
\item
\label{DKL}
L. Dixon, V. Kaplunovsky and J. Louis, Nucl. Phys. {\bf B355} (1991) 649.
\item
\label{orbifold_ref}
L.~Dixon, J.~Harvey, C.~Vafa and E.~Witten,
               Nucl. Phys. {\bf B261} (1985) 678; \\
L.~Dixon, J.~Harvey, C.~Vafa and E.~Witten,
               Nucl. Phys. {\bf B274} (1986) 285; \\
K.~Narain, M.~Sarmadi and C.~Vafa,
               Nucl. Phys. {\bf B288} (1987) 551; \\
K.~Narain, M.~Sarmadi and C.~Vafa,
               Nucl. Phys. {\bf B356} (1991) 163; \\
L.E.~Ib\`a\~nez, H.P.~Nilles and F.~Quevedo,
               Phys. Lett {\bf B187} (1987) 25; \\
L.E.~Ib\`a\~nez, H.P.~Nilles and F.~Quevedo,
               Phys. Lett {\bf B192} (1987) 332.
\item
\label{mod_weights}
S. ~Ferrara, D.~L\"ust and S.~Theisen, Phys. Lett.{\bf B233} (1989) 147; \\
L.E.~Ib\`a\~nez and D.~L\"ust, Nucl. Phys. {\bf B382} (1992) 305.
\end{enumerate}

%-----------------------------------------------------------------------------
\end{document}